\documentclass[11pt]{article}
\usepackage{color,latexsym,amsmath,amssymb,path,url}

\newcommand{\ignore}[1]{}

\newtheorem{theorem}{Theorem}[section]
\newtheorem{lemma}[theorem]{Lemma}
\newtheorem{corollary}[theorem]{Corollary}

\newcommand{\Proof}[1]
        {
        \noindent
        \emph{Proof #1.}~
        }
\newsavebox{\smallProofsym}                     
\savebox{\smallProofsym}                        %
        {
        \begin{picture}(7,7)                    %
        \put(0,0){\framebox(6,6){}}             %
        \put(0,2){\framebox(4,4){}}             %
        \end{picture}                           %
        }                                       %
\newcommand{\smalleop}[1]
        {
        \mbox{} \hfill #1~~\usebox{\smallProofsym}\!\!\!\!\!\!\
        }
\newenvironment{theProof}[1]
        {
        \Proof{#1}}{\smalleop{}
        \medskip

        }
\usepackage{graphicx,psfrag}
\usepackage[rflt]{floatflt}

\newcommand{\placefig}[2]
        {\includegraphics[width=#2]{#1.eps}}
\usepackage{dsfont}

\newcommand{\ex}[1]     {{\,{\ensuremath{\mathbb E}}\!\left\{#1\right\}}}

\newcommand{\pg}[1]{{\cal P}(#1)}
\newcommand{\pgi}[2]{{\cal P}_{#2}(#1,i)}
\newcommand{\pgr}[1]{{\mathsf{pg}}(#1)}
\newcommand{\pgrcc}[2]{{\mathsf{pg}^\le_{#2}}(#1)}
\newcommand{\pgrc}[2]{{\mathsf{pg}^>_{#2}}(#1)}
\newcommand{\pgri}[2]{{\mathsf{pg}}(#1,#2)}
\newcommand{\pgrcci}[3]{{\mathsf{pg}_{#2}^{\le}}(#1,#3)}
\newcommand{\pgrci}[3]{{\mathsf{pg}_{#2}^{>}}(#1,#3)}

\newcommand{\tr}[1]{{\cal T}(#1)}
\newcommand{\tri}[1]{{\mathsf{tr}}(#1)}

\newcommand{\qp}[2]{{\cal Q}_{#1}(#2)}
\newcommand{\qpg}[2]{{\mathsf{qp}_{#1}}(#2)}

\def\hv{{\hat{v}}}

\begin{document}
\pagenumbering{arabic}
\date{}

\title{Counting Plane Graphs: Cross-Graph Charging Schemes\thanks{%
Work on this paper was partially supported by Grant 338/09 from
the Israel Science Fund and by the Israeli Centers of Research Excellence (I-CORE) program (Center  No. 4/11).
Work by Micha Sharir was also
supported by NSF Grant CCF-08-30272,
by Grant 2006/194 from the U.S.-Israel Binational Science Foundation,
and by the Hermann
Minkowski--MINERVA Center for Geometry at Tel Aviv University.
A preliminary version of the paper has appeared at the 20th Internat. Sympos. Graph Drawing (2012).}}

\author{
Micha Sharir\thanks{%
School of Computer Science, Tel Aviv University,
Tel Aviv 69978, Israel and Courant Institute of Mathematical
Sciences, New York University, New York, NY 10012, USA.
{\sl michas@tau.ac.il} }
\and
Adam Sheffer\thanks{%
School of Computer Science, Tel Aviv University,
Tel Aviv 69978, Israel.
{\sl sheffera@tau.ac.il}
}}

\maketitle

\setcounter{footnote}{0}

\begin{abstract}
We study cross-graph charging schemes for graphs drawn in the plane. These are charging schemes where charge is moved across vertices of different graphs.
Such methods have been recently applied to obtain various properties of triangulations that are embedded over a fixed set of points in the plane.
We show how this method can be generalized to obtain results for various other types of graphs that are embedded in the plane.
Specifically, we obtain a new bound of
$O^*\left(187.53^N \right)$ (where the
$O^*(\cdot)$ notation hides polynomial factors) for the maximum number of crossing-free straight-edge graphs that can be embedded over any specific
set of $N$ points in the plane (improving upon the previous best upper bound $207.85^N$ in Hoffmann et al.~\cite{HSSTW11}).
We also derive upper bounds for numbers of several other types of plane graphs (such as connected and bi-connected plane graphs),
and obtain various bounds on expected vertex-degrees in graphs that are uniformly chosen from the set of all crossing-free straight-edge graphs that can be
embedded over a specific point set.

We then show how to apply the cross-graph charging-scheme method for graphs that allow certain types of crossings. Specifically, we consider graphs with no set of $k$ pairwise-crossing
edges (more commonly known as $k$-quasi-planar graphs). For $k=3$ and $k=4$, we prove that, for any set $S$ of $N$ points in the plane,
the number of graphs that have a straight-edge $k$-quasi-planar embedding over $S$ is only exponential in $N$.
\end{abstract}

\section{Introduction}

{\bf Background.}
Consider the following problem --- given a set $S$ of labeled points in the plane, no three collinear, what is the number of graphs that have a straight-edge crossing-free embedding over $S$?
That is, we consider graphs whose vertex set is (or is mapped to) $S$ and whose edges are drawn as straight segments connecting the corresponding pairs of points,
so that these segments do not cross each other (at a point in their relative interiors).
For example, if $S$ is a set of $N$ points in convex position, the answer is known to be $\Theta \left((6+4\sqrt{2})^N\right) \approx \Theta\left(11.66^N\right)$ \cite{FN99}.
The more general problem asks for the maximum number of crossing-free straight-edge graphs that can be embedded over any specific set
of $N$ points in the plane.
The first exponential bound, $10^{13N}$, on the number of such graphs was proved by Ajtai~et al.~\cite{ACNS82} back in 1982.
Since then, progressively (and significantly) smaller upper bounds have been derived (for example, see \cite{HSSTW11,RSW08,SS10}).
Upper bounds on numbers of more specific types of crossing-free straight-edge graphs,
such as Hamiltonian cycles, spanning trees, perfect matchings, and triangulations, were also studied (e.g., see \cite{BKKSS07,BS10,Rib05,Rot05,SW06}).
Worst-case lower bounds for these numbers have also been obtained
(e.g., see \cite{AHHHKV07,DSST11,GNT00}).\footnote{We try to keep a comprehensive list of the various up-to-date bounds in a dedicated webpage
\url{http://www.cs.tau.ac.il/~sheffera/counting/PlaneGraphs.html} (version of August 2012).}

Research on the above problems has led to the development of several useful combinatorial techniques, many of which are interesting in their own right.
One such distant achievement was the introduction of
the \emph{Catalan numbers} by Euler and Lam\'e \cite{Euler61,Lame38}.
A more recent development was the derivation of the \emph{crossing lemma}, obtained by
Ajtai~et al.~\cite{ACNS82}. In this paper we discuss another novel combinatorial technique that has recently emerged from research on the above counting problems.
Namely, this is the concept of \emph{cross-graph charging schemes}.

The idea of applying charging schemes to obtain graph properties probably originated from the attempts of Heesch to prove the \emph{four colors theorem} \cite{Hee69}.
Later, his ideas were used in Appel and Haken's famous proof of the theorem \cite{AH77}, and their extensions have become a common technique in graph theory
(e.g., see \cite{AT07,RT08}).
This technique involves giving charges to vertices (or edges, or faces, for graphs drawn in the plane) of a graph $G$, and then moving these charges between various vertices (or edges, or faces) of $G$.
The novel approach of moving such charges between vertices and edges \emph{of different graphs over the same point set} originated by Sharir and Welzl in 2006 \cite{SW06},
in studying the maximum number of triangulations that can be embedded over a specific set of $N$ points in the plane.
Since then, this technique has been extended in \cite{RSW08,RW09,SS10,SSW10}
to study various combinatorial and algorithmic properties of triangulations.

In this paper, we extend the idea of cross-graph charging schemes beyond the realm of triangulations.
We first show how to apply this technique to bound the maximum number of crossing-free straight-edge graphs that can be embedded over a specific set of points in the plane.
Then we show how to extend this idea to several other types of graphs, including families of non-planar graphs (this seems to be the first derivation of reasonable bounds for such graph types).
It seems likely that these techniques can be further extended to other types of problems
(that is, problems not involving bounding or counting the number of embedded graphs), and we hope
that the present study will motivate such applications. Before discussing our results any further, we require some formal definitions of the concepts related to these problems.

{\bf Notations and results.}
A {\em planar graph} is a graph that can be embedded in the plane in such
a way that its vertices are embedded as points and its edges are embedded as Jordan arcs that connect the respective pairs of points and can meet only at a common endpoint.
A {\em crossing-free straight-edge graph} is a plane embedding of a planar graph
such that its edges are embedded as non-crossing straight line
segments; we sometimes refer to such graphs simply as \emph{plane graphs}.
In Section~\ref{sec:plane} we only consider plane graphs.
In Section~\ref{sec:quasi} we allow certain types of crossings by considering \emph{quasi-planar graphs}; here too we assume that the edges are embedded as (possibly crossing) straight line segments.
In both sections we only consider embeddings where the points are in general position, that is,
where no three points are collinear. For upper bounds on the number of graphs, this involves no loss of generality, because the number of graphs can only grow
when a degenerate point set is slightly perturbed into general position.

A \emph{triangulation} of a finite point set $S$ in the plane is a
maximal plane graph on $S$ (that is, no additional straight edges can be inserted without crossing any of the existing edges),
so all bounded faces of the planar map that it defines are triangles.
For a set $S$ of points in the plane, we denote by $\tr{S}$ the set of all
triangulations of $S$, and put $\tri{S}:= \left|\tr{S}\right|$. Similarly, we denote by $\pg{S}$ the set of all plane graphs of $S$, and put $\pgr{S}:= \left|\pg{S}\right|$.
Finally, let $\tri{N}=\max_{|S|=N}\tri{S}$ and $\pgr{N}=\max_{|S|=N}\pgr{S}$. So another way of formulating our problem is --- find a small constant $b$
(ideally, find the smallest) such that $\pgr{N}=O^*(b^N)$.\footnote{In the notations $O^*()$, $\Theta^*()$, and $\Omega^*()$, we neglect polynomial factors.} (By the results mentioned above, we know that such a $b$ exists.)

Notice that every plane graph is contained in at least one triangulation.
Also, by Euler's formula, every triangulation has fewer than $3|S|$ edges, and thus, every triangulation contains fewer than $2^{3|S|}=8^{|S|}$ plane graphs.
From the above we have the inequality
$\pgr{S} < 8^{|S|} \cdot \tri{S}$, which implies $\pgr{N} < 8^{N} \cdot \tri{N}$. Every several years an improved upper bound for $\tri{N}$ is discovered (e.g., see \cite{DeSo97,SaSe03,SW06}), and currently, the best known bound is $\tri{N}<30^N$ \cite{SS10}.
Combining this bound with the above inequality implies $\pgr{N} < 240^N$.
Currently, the best known lower bound is  $\pgr{N} = \Omega(41.18^N)$ \cite{AHHHKV07}.

The inequality $\pgr{N} < 8^{N} \cdot \tri{N}$ seems rather weak, since it potentially counts some plane graphs many times
(once for every triangulation containing the graph).
Razen, Snoeyink, and Welzl \cite{RSW08} were the first to address this inefficiency, deriving the slightly improved inequality $\pgr{N} = O\left(7.9792^N\right) \cdot \tri{N}$.
A more significant improvement of $\pgr{N} < 6.9283^N \cdot \tri{N}$ was recently obtained by Hoffmann et al.\ \cite{HSSTW11}.
This implies the bound $\pgr{N} < 207.85^N$.

As far as we know, our cross-graph charging-scheme method is currently the only method
that does not rely on the ratio between $\pgr{N}$ and $\tri{N}$ and yields a non-astronomical bound. An initial, more direct application of this
method implies only a bound of $3207.42^N$. On the other hand, by combining this method with the current bound on the number of triangulations
(indirectly, by using an upper bound on the maximum number of plane graphs with at least $cN$ edges, which is derived in \cite{HSSTW11} and relies on $\tri{N}$)
we obtain $\pgr{N}=O^*\left(187.53^N \right)$.

Our method relies on charging schemes between objects from different plane graphs over the same point set (hence the name \emph{cross-graph charging schemes}).
Given a set $S$ of $N$ points in the plane, we consider the set $S \times \pg{S}$ and
call each of its elements a {\em ving} \ ({\em v}ertex {\em in}
{\em g}raph, similar to the definition of a \emph{vint}---\emph{v}ertex \emph{in t}riangulation---from \cite{SS10,SSW10,SW06}).
Intuitively, a ving is an instance of a vertex (a point of $S$)
in a specific plane graph. Our charging schemes are between vings from different graphs (sharing a common vertex).

A \emph{$k$-quasi-plane} graph is a straight-edge graph over a set of points in the plane that may contain crossings,
but does not contain any set of $k$ pairwise crossing edges (some other works, such as \cite{AT07}, refer to such graphs as $k$-\emph{quasi-planar geometric graphs}).
Notice that a 2-quasi-plane graph is simply a plane graph.

For a set $S$ of points in the plane, we denote by $\qp{k}{S}$ the set of all
$k$-quasi-plane graphs on $S$, and put $\qpg{k}{S}:= \left|\qp{k}{S}\right|$.
Moreover, we let $\qpg{k}{N}=\max_{|S|=N}\qpg{k}{S}$.
As far as we know, there are no previously known singly exponential upper bounds on $\qpg{k}{N}$, for any $k \ge 3$.
We show that an appropriate extension of our technique easily implies the bounds $\qpg{3}{N} \le 2^{26N}$ and $\qpg{4}{N} \le 2^{145N}$.
These bounds are probably very far from being tight, but our purpose here is to show that the number of 3-quasi-plane graphs, say, that can be embedded over a specific point set
is only exponential in the number of points. As a side remark, we note that the first bound, on $\qpg{3}{N}$, is significantly
smaller than the first exponential bound ($10^{13N}$) that was obtained for the number of plane graphs  \cite{ACNS82}.
We also show that the main conjecture about quasi-planar graphs (namely, that the number of their edges is linear for any fixed $k$, e.g., see \cite{Ack09,AT07,pach04}) would imply, if true, that $\qpg{k}{N}$ is (only) exponential
in $N$ for any fixed $k$.
In fact, our exponential bounds for $\qpg{3}{N}$ and $\qpg{4}{N}$ derive from the fact that the number of
edges of $3$-quasi-plane and $4$-quasi-plane graphs is linear in the number of vertices~\cite{Ack09,AT07}.

\section{An Upper Bound on the Number of Plane Graphs} \label{sec:plane}
In this section we derive upper bounds on the number of plane graphs.
In Section~\ref{ssec:direct}, we derive the initial bound $\pgr{N} \le 4096^N$.
In Section~\ref{ssec:firstImp}, we exploit some geometric aspects of the problem, to improve the bound to $\pgr{N} \le 3207.42^N$.
Even though this is far worse than the recent bound $\pgr{N} < 207.85^N$ \cite{HSSTW11},
it constitutes a significant progress in deriving bounds that do not depend on $\tri{N}$.
In Section~\ref{ssec:imp}, we extend our technique to obtain the bound $\pgr{N}=O^*\left(187.53^N \right)$, which is currently the best known upper bound for this quantity.
This extension is a combination of our technique with some recently obtained bounds on the number of certain types of plane graphs (from \cite{HSSTW11}).
These latter bounds do depend on the number of triangulations, but the way we exploit these bounds makes our new bound (that is, $O^*\left(187.53^N \right)$) depend \emph{non-linearly} on $\tri{N}$; see below for details.
In Section~\ref{ssec:connected}, we apply the technique to upper bound the numbers of some other types of plane graphs, such as connected and bi-connected plane graphs.
These bounds are only slightly better than our bound for the number of plane graphs, but it is the first time where these subfamilies admit better upper bounds than for the overall number of plane graphs.
We also show how to obtain various degree-related bounds for vertices in a random plane graph over a fixed set of points, using the same technique.

\subsection{The infrastructure and an initial bound} \label{ssec:direct}
Given two vertices $p,q$ of a plane graph $G$, we say that $p$ \emph{sees} $q$ in $G$
if the (straight) edge $pq$ does not cross any edge of $G$.
The \emph{degree} of a ving $(p,G)$ is the
degree (number of neighbors) of $p$ in $G$; a ving of degree $i$
is called an \emph{$i$-ving}.
We say that a ving $v=(p,G)$ is an \emph{$x$-ving} if we cannot increase the degree of $p$ by inserting additional (straight) edges to $G$
(that is, $p$ cannot see any vertex that is not connected to it in $G$).
We say that a ving $u=(p,G')$ \emph{corresponds} to the $x$-ving $v=(p,G)$ if $G$ is obtained by inserting into $G'$ all the edges that connect $p$ to the points that it sees in $G'$ and is not connected to them.
Notice that every ving corresponds to a unique $x$-ving.
Given a plane graph $G \in \pg{S}$, we denote by $v_i(G)$ the number of $i$-vings in $G$, for $i\ge0$, and by $v_x(G)$ the number of $x$-vings in $G$.
Finally, the expected value of $v_x(G)$, for a graph chosen uniformly at random from $\pg{S}$, is denoted as $\hv_x(S)$. More formally,
$\displaystyle \hv_x = \hv_x(S) := \ex{v_x(G)} = \frac{\sum_{G \in \pg{S}}v_x(G)}{\pgr{S}}$.
A similar notation, $\hv_i(S)$, applies to the expected value of $v_i(G)$.

The following lemma, inspired by similar lemmas in \cite{SS10,SSW10,SW06}, presents a connection between $\hv_x$ and upper bounds for $\pgr{N}$.

\begin{lemma} \label{le:NumberOfPG}
For $N \geq 2$, let $\delta_{N}>0$ be a real number, such that
$\hat{v}_x(S) \geq \delta_{N} N$ holds for every set $S$ of $N$
points in the plane in general position. Then
$\displaystyle \pgr{N} \leq \frac{1}{\delta_N} \pgr{N-1}$.
\end{lemma}
\begin{theProof}{\!\!}
Let $S$ be a set that maximizes $\pgr{S}$
among all sets of $N$ points in the plane.
Note that we can get some plane graphs of $S$ by
choosing a point $q \in S$ and
a plane graph $G$ of $S \setminus \{q\}$, inserting $q$ into $G$, and then connecting $q$ to all of the vertices that it can see in $G$.
In fact, a plane graph $G$ of $S$ can be obtained
in exactly $v_x(G)$ ways in this manner (in particular,
if $v_x(G) = 0$, $G$ cannot be obtained at all in this fashion).
This is easily seen to imply that
\[
\hv_x\cdot \pgr{S} = \mbox{$\sum_{G\in \pg{S} }$} v_x(G) = \mbox{$\sum_{q \in S}$} \pgr{S \setminus\{q\}}\enspace.
\]

The leftmost expression equals $\hv_x\cdot \pgr{N}$,
and the rightmost expression is at most $N\cdot \pgr{N-1}$. Hence, with $\hv_x \geq \delta_{N} \, N$, we have
$\displaystyle \pgr{N} = \pgr{S} \leq \frac{N}{\hv_x} \cdot \pgr{N-1} \leq
\frac{1}{\delta_{N}} \cdot \pgr{N-1}$. \vspace{-5mm}

\end{theProof}

We thus seek a lower bound for $\hv_x$, of the kind assumed in Lemma \ref{le:NumberOfPG}. For this purpose, we use a charging scheme similar in spirit to
the one presented in \cite{SS10,SSW10,SW06}.
The following lemma establishes such a bound, which is rather weak. Nevertheless, it has the advantage of being a ``stand-alone" bound, independent of
(bounds on) the
number of triangulations on $S$. In the following subsections, we will derive a considerably improved bound, which does depend on
(the best known bound on) the number of triangulations of $S$
(albeit in a nonlinear manner).

\begin{lemma}\label{le:LowerV0}
For every set $S$ of $N$ points in the plane in general position, $\displaystyle \hv_x(S) \ge \frac{N}{4096}$.
\end{lemma}
\begin{theProof}{\!\!}
We use a charging scheme where every $i$-ving  $v=(p,G)$ is given $7-i$ units of charge.
The sum of the charges of the vings in any fixed plane graph $G \in \pg{S}$ is $\sum_i(7-i)v_i(G) = 7\sum_iv_i(G) - \sum_{i}iv_i(G) = 7N - \sum_{i}iv_i(G)$.
Since $G$ can have at most $3N-6$ edges, we have $\sum_{i}iv_i(G) \le 6N-12$.
This implies that the total charge in any fixed graph is at least $7N - \sum_{i}iv_i(G) \ge N+12$. Therefore, on average, every ving has a charge larger than 1.

Next, every $i$-ving moves its entire charge to its corresponding $x$-ving.
(In general, the $x$-ving lies in a graph different than the one containing the $i$-ving.)
This results with all of the charge being placed only on $x$-vings.
If we can show that every $x$-ving gets charged at most $t$ units in this manner, we will get the lower bound $\hv_x \ge N/t$, as is easily verified.

To upper bound the charge that an $x$-ving $v=(p,G)$ can get, we need to consider the degree $d$ of $v$.
Notice that the number of $i$-vings that charge $v$ is exactly $\binom{d}{i}$ (that is, the number of ways to remove $d-i$ edges that are adjacent to $p$ in $G$).
Therefore, the total charge to $v$ is
\[\sum_{0 \le i \le d}\binom{d}{i}(7-i) = 7\sum_{0 \le i \le d}\binom{d}{i}-\sum_{0 \le i \le d}\binom{d}{i}i = 7\cdot 2^d - d\cdot 2^{d-1} = 2^{d-1}(14-d).\]

\noindent This expression maximizes when $d$ is either 12 or 13, and is then 4096.
Thus, on average, a plane graph of $S$ has more than $\frac{N}{4096}$ $x$-vings.
\end{theProof}

Combining Lemmas \ref{le:NumberOfPG} and \ref{le:LowerV0} and using an obvious induction on $N$
(starting with $\pgr{1}=1$), we obtain
\begin{theorem}
$\displaystyle \pgr{N} \le 4096^N$.
\end{theorem}
{\bf Remark.} The above analysis remains valid if we charge 0-vings instead of $x$-vings.
That is, each $i$-ving $(p,G)$ passes its charge to the 0-ving $(p,G')$ obtained by removing all the edges incident to $p$ in $G$.
This is also the case for several of the other proofs in this paper, but not for all of them (several proofs in
Section~\ref{ssec:connected} do not seem to be amenable to this modification).

\subsection{First improvement} \label{ssec:firstImp}
Interestingly, the bound in Section~\ref{ssec:direct} hardly relies on the geometric properties of the problem.
Specifically, it only uses Euler's formula for plane graphs\footnote{In fact, it only uses the fact that the number of edges in a plane graph is at most three times the number of vertices.}, and the trivial property, already noted, that in a plane graph, connecting a ving to any subset of the vertices that it sees results in a (larger) plane graph.
In this subsection we obtain an improved bound by observing and exploiting some additional geometric properties of $x$-vings.

\begin{lemma}\label{le:LowerV0imp}
For every set $S$ of $N$ points in the plane in general position, $\displaystyle \hv_x(S) \ge \frac{N}{3207.42}$.
\end{lemma}

\begin{theProof}{\!\!}
We start by applying the same charging scheme as in the proof of Lemma \ref{le:LowerV0}, but
then perform another step of moving charges across  $x$-vings, as follows.
We say that an $x$-ving $v=(p,G)$ is an $x_i$-ving if $v$ is also an $i$-ving.
According to the analysis in the proof of Lemma \ref{le:LowerV0}, only $x_{12}$-vings and $x_{13}$-vings are charged 4096;
the next highest charge (for $x_{11}$-vings) is 3072.
At the other end, an $x_{3}$-ving is charged only 44, and an $x_2$-ving is charged only 24.
Note that an $x$-ving $(p,G)$ can be an $x_2$-ving only if $p$ is part of the boundary of the convex hull of $S$ and the two neighbors of $p$ along this boundary are connected in $G$
(e.g., see Fig.~\ref{fi:0vings}(a)). Note also that $x_1$-vings (and $x_0$-vings) do not exist.
Consider an $x_{i}$-ving $v=(p,G)$, where $i > 3$, and let $S_v$ be the set of $i$ vertices
that are connected to $p$ in $G$.
Let $P_v$ be the star-shaped polygon (with respect to $p$) that is obtained by removing from $G$ all the edges that are incident to $p$, ordering the vertices of $S_v$ in their angular (cyclic) order around $p$, and connecting every pair of consecutive vertices by an edge
(some of these connecting edges may already exist in $G$, and adding the others cannot create a crossing, because $v$ is an $x$-ving).
Triangulate $P_v$ arbitrarily and let $\Delta$ denote the triangle that contains $p$.
We remove from $G$ all the edges incident to $v$, add the edges of $\Delta$, connect $p$ to the three vertices of $\Delta$
to obtain a new graph $G'$, and notice that $v'=(p,G')$ is an $x_3$-ving (once again, some of the edges of $\Delta$, but not all of them, may already exist in $G$).
Notice that we did not add the missing edges of $P_v$ to $G'$.
We say that $v$ is \emph{reduced} to $v'$.
An example for such a reduction is depicted in Fig.~\ref{fi:0vings}(b).

\begin{figure}[t]

\centerline{\placefig{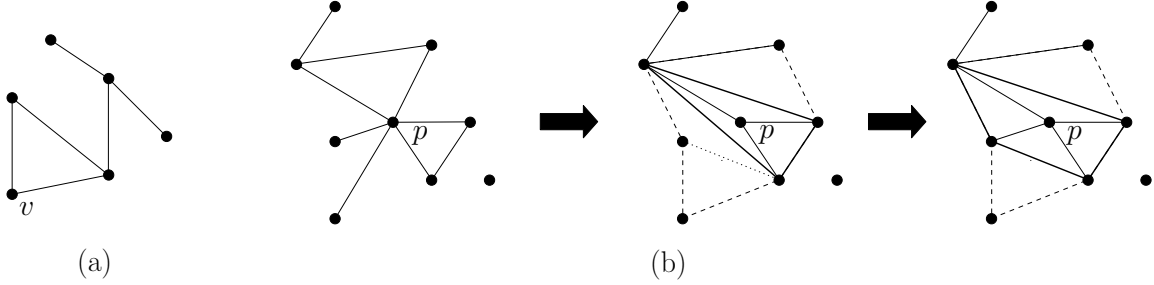}{1\textwidth}}
\vspace{-4mm}

\caption{\small \sf (a) The ving involving $v$ is an $x_2$-ving.
(b) An $x_6$-ving $v=(p,G)$ and an $x_3$-ving it reduces to: The enclosing polygon $P_v$ (whose new edges are drawn dashed), and a triangulation of $P_v$, with the
triangle containing $p$ highlighted; only the solid edges belong to the new graph $G'$.
A corresponding $x_4$-ving to which we can reduce $v$ is also depicted.
}
\label{fi:0vings}
\vspace{-2mm}
\end{figure}
Given a specific $x_{3}$-ving $v=(p,G)$, we now consider how many $x_{12}$-vings and $x_{13}$-vings can be reduced to it.
Let $\Delta$ denote the triangle spanned by the three vertices $u,v,w$ that $p$ is connected to. (By construction, only $x_3$-vings where all the edges of $\Delta$ belong to $G$ should be considered.)
Denote by $a,b,c$
the number of additional vertices that $p$ would be able to see after the removal of each of the three respective edges
(say, $uv$, $vw$, $wu$) of $\Delta$ in $G$.
For example, if we remove the edge $uv$ of $\Delta$ then $p$ would see $3+a$ vertices (including the three vertices of $\Delta$ that are connected to $p$),
if we remove all three edges of $\Delta$ then $p$ would be able to see $3+a+b+c$ vertices, and so on.
After such an edge removal, we can connect $p$ to all of the new vertices that it sees, and obtain an $x$-ving that reduces to $v$.
(Every $x$-ving that reduces to $v$ is obtained in this manner.)
For every set of values of $a,b,c$, out of the seven possible edge removal combinations, at most four could yield an $x_{12}$-ving or an $x_{13}$-ving.
That is, out of the seven numbers $3+a$, $3+b$, $3+c$, $3+a+b$, $3+a+c$, $3+b+c$, $3+a+b+c$, at most four can be equal to $12$ or $13$.
For example, four combinations are obtained when $a=9$, $b=1$, and $c=0$.  This can be verified by a simple case analysis,
depending on how many of $a,b,c$ are equal to 9 or 10 (so that the corresponding quantity $3+a$, $3+b$, or $3+c$ is 12 or 13).
Thus, at most four $x_{12}$-vings and $x_{13}$-vings can reduce to any specific $x_{3}$-ving. From every $x_{12}$-ving and every $x_{13}$-ving, we move
a charge of 810.4 to some $x_3$-ving that it reduces to. Now, every $x$-ving is charged at most $3285.6 = 44 + 4\cdot 810.4 = 4096 - 810.4$ ($x_{i}$-vings
are charged at most 0 when $i>13$, at most 3072 when $3<i<12$, and 24 when $i=2$).
This already gives us the bound $\hv_x(S) \ge N/3285.6$.

We can further improve this bound by also considering $x_{4}$-vings.
Consider an $x_{i}$-ving $w=(p,G)$ where $i=12$ or 13, the triangulated polygon $P_w$, and the respective
$x_{3}$-ving $u=(p,H)$ that lies inside a triangle $\Delta$ of $P_w$. Consider some triangle $\Delta'$ that is adjacent to $\Delta$ in $P_w$ (there always exists at least one such triangle).
Insert the edges of $\Delta'$ into $H$ (those that do not already lie in $H$), remove the edge that is incident to both triangles, and connect $p$ to the fourth vertex of the resulting quadrilateral (recall that $P_w$ is star-shaped with respect to $p$), to obtain a new graph $H'$.
Notice that the $x$-ving $u'=(p,H')$ is connected to four vertices in $H'$, and is thus charged 80 (by the scheme in the proof of Lemma \ref{le:LowerV0});
an example is depicted in Fig.~\ref{fi:0vings}(b).

Consider how many $x_3$-vings may lead to an $x_4$-ving $u'=(p,H')$ in the manner just described.
Given $u'$, there are at most two ways to choose a diagonal $d$ of the quadrilateral containing $p$ in $H'$.
Inserting $d$ into $H'$ (and removing the edge that crosses $d$) divides this quadrilateral into two triangles. Let $\Delta'$ be the triangle that does not contain $p$, and
let $e$ and $e'$ be the two edges of $\Delta'$ that are contained in $H'$ (i.e., different from $d$).
Both $e$ and $e'$ may or may not belong to the graph of the $x_3$-ving, and combining this with the choice of $d$ implies that there are at most eight $x_{3}$-vings that lead to $u$ in the manner just described.
Each of these eight $x_{3}$-vings is charged by at most four $x_{12}$- and/or $x_{13}$-vings, as described earlier. We move a charge of 78.18 from each of these at most forty $x$-vings (eight $x_3$-vings and 32 $x_{12}$- and/or $x_{13}$-vings) to $u'$.
Then, no $x$-ving is charged more than $3285.6-78.18 = 3207.42 > 80+40\cdot 78.18$.
\end{theProof}
{\bf Remark:} One could obtain a better bound by also considering $x_{5}$-vings, $x_{6}$-vings, etc.
Since this seems to imply only a slight improvement and requires a somewhat tedious analysis, we do not go into further details in this paper,
especially since we are after a much more drastic improvement, given in the next subsection.
The technique used in this subsection will be used in the following one to gain an additional improvement in the bound derived there.

By combining Lemmas \ref{le:NumberOfPG} and \ref{le:LowerV0imp}, we obtain our second upper bound:
\begin{theorem}
$\displaystyle \pgr{N} \le 3207.42^N$.
\end{theorem}

\subsection{Second improvement} \label{ssec:imp}

Given a set $S$ of $N$ points in the plane, we let $\pgrc{S}{c}$ (resp., $\pgrcc{S}{c}$) denote the number of plane graphs with more than $cN$ edges
(resp., at most $cN$ edges) that can be embedded over $S$, for some parameter $0< c < 3$.
Additionally, let $\hv_{x,m}(S)$ denote the expected (i.e., average) value of $v_x(G)$ over all plane
graphs $G \in \pg{S}$ with at most $m$ edges.

In \cite{HSSTW11}, Hoffmann et al.~establish the following theorem:
\begin{theorem}
\label{th:triVSpgcc}
For any set $S$ of $N$ points in the plane and $19/12 \le c \le 3$,
\[ \pgrc{S}{c} = O^*\left( \left( \frac{5^{5/2}}{8\left(c+t-\frac{1}{2}\right)^{c+t-\frac{1}{2}} (3-c-t)^{3-c-t}(2t)^t\left(\frac{1}{2}-t\right)^{\frac{1}{2}-t}} \right)^N \hspace{-2mm} \tri{S} \right), \]
where
$\displaystyle t = \frac{1}{2} \left(\sqrt{(7/2)^2 + 3c + c^2 } - 5/2 - c \right)$.
\end{theorem}
We begin by stating the following variants of Lemmas \ref{le:NumberOfPG} and \ref{le:LowerV0}.
\begin{lemma}
\label{le:NumberOfPG2}
Let $S$ be a set of $N$ points in the plane and let $0 < c < 3$ be a parameter, such that $\hv_{x,cN}(S) \geq \delta N$  for some constant $\delta >0$. Then
$\displaystyle \pgrcc{S}{c} \leq \frac{1}{\delta}\cdot\pgr{N-1}$.
\end{lemma}
\begin{theProof} {\!\!}
By applying the same proof as in Lemma \ref{le:NumberOfPG}, we obtain the relation
$\displaystyle \pgrcc{S}{c}\le (1/\delta) \cdot \pgrcc{N-1}{cN/(N-1)}$.
The lemma follows by noting that $\pgrcc{N-1}{cN/(N-1)} \le \pgr{N-1}$.
The reason for replacing $c$ by $cN/(N-1)$ is that the graphs obtained by removing $x$-vings have only $N-1$ vertices
(and fewer than $cN$ edges). \end{theProof}
\vspace{1 mm}
We let $c = 1.968549$; see a remark below that explains this choice.
Substituting this value of $c$ into Theorem \ref{th:triVSpgcc}, and using $\tri{N} < 30^N$ from \cite{SS10}, we get
$\pgrc{N}{c} = O^*\left(187.53^N\right)$.

\begin{lemma}
\label{le:LowerV02}
For every point set $S$ of $N$ points in the plane in general position,
either $\displaystyle \pgr{S} = O^*(187.53^N)$ or $\displaystyle \hv_{x,cN}(S) > N/187.53$ (or both).
\end{lemma}
\begin{theProof}{\!\!}
We first assume that $N$ is at least some sufficiently large constant $N_0$, whose choice is dictated by the forthcoming analysis. For $N < N_0$ the first bound in the lemma holds trivially, for an appropriate choice of the constant of proportionality.

We use a charging scheme in which an $i$-ving $v=(p,G)$ is given $a+2m/N-i$ units of charge, where $c = 1.968549$ as above,
$a=(9-4c)/2$, and $m$ is the number of edges in $G$.
The sum of the charges of the vings in any fixed plane graph $G \in \pg{S}$ with $m$ edges is
\[\sum_i\left(a+2m/N-i\right)v_i = (a+2m/N)\sum_i v_i - \sum_i{i}v_i = aN + 2m - \sum_{i}iv_i. \]
Since $G$ has $m$ edges, we have $\sum_{i}iv_i = 2m$.
This implies that the charge in any fixed graph of this kind is $aN$, and that, on average, every ving has a charge of $a$.
Moreover, the total charge over all of the vings is

\begin{equation} \label{eq:lowerC}
C = aN \cdot \pgr{S} = aN \cdot \left(\pgrc{S}{c} + \pgrcc{S}{c}\right)\enspace.
\end{equation}

Next, we move the charge to $x$-vings in exactly the same manner as in Lemma \ref{le:LowerV0}.
That is, every ving moves its entire charge to its corresponding $x$-ving.
Consider an $x_d$-ving $v=(p,G)$.
The number of $i$-vings that charge $v$ is exactly $\binom{d}{i}$. Each of these $i$-vings holds a charge of $a+2(m-d+i)/N-i$
(because it belongs to a graph with $m-d+i$ edges).
Therefore, the total charge made to $v$ is

\[ \sum_{0 \le i \le d}\binom{d}{i}\left(a+\frac{2(m-d+i)}{N}-i\right)  <
         \left(a+\frac{2m}{N}\right) \cdot\sum_{0 \le i \le d}\binom{d}{i}- \sum_{0 \le i \le d}\binom{d}{i}i  \]
\vspace{-3mm}

\[ = \left(a+\frac{2m}{N}\right) \cdot 2^d - d\cdot 2^{d-1}  = 2^{d-1}\left(2a+\frac{4m}{N}-d\right)\enspace. \]

Since $m<3N$, this expression maximizes when $m\approx 3N$ and $d=13$, and is then $\nu <5144.58$.
In what follows we assume that $m\le cN$, because the number of graphs with $m>cN$ is at most $\pgrc{N}{c} = O^*\left(187.53^N\right)$.
For $m\le cN$ the expression is smaller than $2^{d-1}(9-d)$, which maximizes when $d$ is either 7 or 8,
and is then 128.

We again move the charge, this time in the manner described in Section~\ref{ssec:firstImp}.
Before moving charges, $x_7$-vings and $x_8$-vings are charged fewer than 128 units,
$x_6$-vings are charged fewer than 96, $x_5$-vings are charged fewer than 64,
$x_4$-vings fewer than 40, $x_3$-vings fewer than $24$, and $x_2$-vings are charged fewer than 14.
It is easily checked that, as before, at most four $x_7$-vings and $x_8$-vings can reduce to a single $x_3$-ving.
After moving a charge of 20.8 from each $x_7$- and $x_8$-ving to the $x_3$-ving that it reduces to, all of these $x$-vings have a charge smaller than
$107.2 = 128 -20.8 = 24 + 4\cdot20.8$.
Continuing as in Section~\ref{ssec:firstImp}, we now move to every $x_4$-ving a charge of 1.639 from at most forty $x$-vings (specifically, from $x_7$-vings, $x_8$-vings,
 and $x_3$-vings that took part in the previous exchange).
After this step, every $x$-ving has a charge smaller than $105.561 = 107.2 - 1.639 > 40 + 40\cdot1.639$
(as noted, every $x_i$-ving, for $i\ne 3,7,8$, is charged at most 96).
Denoting by $\mu < 105.561$ the maximum modified charge of an $x$-ving, we have
\begin{equation} \label{eq:upperC}
C\le \mu \hspace{-4mm} \sum_{|E(G)| \le cN} \hspace{-3.5mm} v_x(G) + \nu \hspace{-4mm} \sum_{|E(G)| \ge cN} \hspace{-3.5mm} v_x(G) \le
\mu \hspace{-4mm} \sum_{|E(G)| \le cN} \hspace{-3.5mm} v_x(G) + \nu \cdot N \cdot \pgrc{S}{c}\enspace.
\end{equation}

By combining \eqref{eq:lowerC} and \eqref{eq:upperC}, we get
\[ aN \cdot (\pgrc{S}{c} + \pgrcc{S}{c}) \le \mu \hspace{-4mm} \sum_{|E(G)| \le cN} \hspace{-3.5mm} v_x(G) + \nu \cdot N \cdot \pgrc{S}{c}\enspace. \]

Isolating the term $\mu \sum_{|E(G)| \le cN} v_x(G)$ and dividing by $\pgrcc{S}{c}$ yields
\begin{equation} \label{eq:doubleCount}
\mu \cdot \hv_{x,cN}(S) = \frac{\mu\sum_{|E(G)| \le cN} v_x(G)}{\pgrcc{S}{c}} \ge aN - \frac{\pgrc{S}{c}}{\pgrcc{S}{c}} N \cdot (\nu-a)\enspace.
\end{equation}
Consider some sufficiently fast increasing polynomial $\phi(N)$ (e.g., $\phi(N)= N^{10}$). If $\pgrcc{S}{c} \le \phi(N) \cdot \pgrc{S}{c}$
then by Theorem \ref{th:triVSpgcc} we have $\pgr{S} = \pgrcc{S}{c} + \pgrc{S}{c} = O^*(\pgrc{S}{c}) = O^*(187.53^N)$, and the lemma follows.
On the other hand, if $\pgrcc{S}{c} > \phi(N) \cdot \pgrc{S}{c}$ then the absolute value of the rightmost term in \eqref{eq:doubleCount}
can be upper bounded by $\varepsilon=(\nu-a)/N^9$, and (\ref{eq:doubleCount}) then implies, for $N\ge N_0$,

\[\hv_{x,cN}(S) \ge \frac{aN -\varepsilon}{\mu} \ge
\frac{aN -\frac{\nu - a}{N_0^9}}{\mu}
> \frac{(9-4c)N-\frac{\nu-a}{N_0^9}}{211.122} > N/187.53\enspace, \]
where the last inequality holds when $N_0$ is sufficiently large.
\end{theProof}

By combining Lemmas \ref{le:NumberOfPG2} and \ref{le:LowerV02}, we get the following improved bound.
\begin{theorem} \label{th:pg2}
$\displaystyle \pgr{N} = O^*\left(187.53^N\right)$.
\end{theorem}
\begin{theProof}{\!\!}
Let $S$ be a set of $N$ points in the plane that maximizes $\pgr{S}$ (that is, $\pgr{S} = \pgr{N})$.
As mentioned above,
 $\displaystyle \pgrc{S}{c} \le \pgrc{N}{c} = O^*\left(187.53^N\right)$.
\noindent Hence,
\begin{equation}
\pgr{N} = \pgr{S} = \pgrcc{S}{c} + \pgrc{S}{c} \le \pgrcc{S}{c} + O^*\left(187.53^N\right)\enspace.
\end{equation}

\noindent By Lemma \ref{le:LowerV02} we have either $\pgrcc{S}{c} = O^*\left(187.53^N\right)$ or $\hv_{x,cN}(S) > N/187.53$. The former case immediately implies the asserted bound, and in the latter case we have, by Lemma \ref{le:NumberOfPG2},
\[ \pgr{N} \le 187.53\cdot\pgr{N-1} + O^*\left(187.53^N\right)\enspace, \]
and the asserted bound follows by induction on $N$.
\end{theProof}

\noindent {\bf Remarks:} (1) The bound in Theorem \ref{th:pg2} can be slightly improved by passing some of the charge to $x_5$-vings and $x_6$-vings.

\noindent (2) Here is an explanation for our choice of $c$. It seems to yield the best
bound on $\pgr{N}$, although we have no formal proof of this.
Informally, we aim at a situation where the choice of $c$ and $a$
has the property that there exists $k$ such that
$x_k$-vings and $x_{k+1}$-vings have the same bound on their maximum charge.
We then pass some of the charge to $x_3$-vings and later to $x_4$-vings,
so that the new charges of $x_3$-vings, $x_4$-vings, $x_k$-vings,
and $x_{k+1}$-vings are all bounded by the same quantity, which is still
larger than the bounds on the charges of all the other $x_i$-vings,
charges that have not been touched by this charge-moving process.

The $k$ that we have chosen was $k=7$. The bound on the
original charge of $x_7$-vings is $2^6(2a+4c-7)$, and that of
$x_8$-vings is $2^7(2a+4c-8)$. To make them equal, we have to choose
$2a+4c=9$, as we did. We now proceed as described above, moving
$20.8$ units of charge in the first step, from every $x_7$-ving
and every $x_8$-ving to the $x_3$-vings they reduce to, and then
moving $1.639$ units to $x_4$-vings from $x_3$-vings, $x_7$-vings,
and $x_8$-vings, as prescribed above. Note that these numbers do not
depend on the choice of $c$ and $a$, only on the property that
$2a+4c=9$. We thus get the maximum modified charge
$\mu < 105.561$, and, as the analysis shows, the base of the exponential
bound is the maximum of $\mu/a = 2\mu/(9-4c)$ and $\tau\beta(c)$,
where $\beta(c)$ is the base of the exponential bound given in
Theorem~\ref{th:triVSpgcc},
and $\tau=30$ is the base in the best known bound on the number of
triangulations.

In other words, we need to find $c$ that balances between
$2\mu/(9-4c)$ and $\tau\beta(c)$, which we have done
using the Wolfram Mathematica software,\footnote{Wolfram Research, Inc.,
Mathematica, Version 7.0.1, Champaign, IL, (2009).}
which has produced the optimal value $c \approx 1.968549$.

We note that, as
opposed to previous bounds for $\pgr{N}$, the dependence of the new bound on $\tri{N}$ is non-linear.

\subsection{Additional types of plane graphs and degree-related bounds} \label{ssec:connected}
In this subsection we present various additional bounds that can be obtained by using the above technique.
Specifically, we extend the technique to some other types of plane graphs,
and show how to derive degree-related properties of random plane graphs (embedded over a fixed set $S$).

Given a set $S$ of $N$ points in the plane, we let $\pgri{S}{i}$ denote the number of plane graphs that can be embedded over $S$ and that contain no vertex of degree smaller than $i$.
We let $\pgrci{S}{c}{i}$ (resp., $\pgrcci{S}{c}{i}$) denote the number of plane graphs with more than $cN$ edges
(resp., at most $cN$ edges) and with no vertex of degree smaller than $i$, that can be embedded over $S$.
Additionally, let $\hv_{x,m}(S,i)$ denote the expected number of $x$-vings in a graph uniformly chosen from the set of
graphs that are in $\pg{S}$, have at most $m$ edges, and contain no vertex of degree smaller than $i$.

\begin{lemma}
\label{le:NumberOfPGno0}
Let $S$ be a set of $N$ points in the plane such that $\hv_{x,cN}(S,i) \geq \delta N$ for an integer $i \ge 0$ and some parameters $0<c<3$ and $\delta>0$. Then
\[\pgrcci{S}{c}{i} \leq \frac{1}{\delta}\cdot\pgr{N-1}\enspace.\]
\end{lemma}
\begin{theProof} {\!\!}
Denote by $\pgi{S}{m}$ the set of plane graph with at most $m$ edges
and with no vertex of degree smaller than $i$, that can be embedded over $S$.
The proof goes along the same lines of the proofs of Lemmas \ref{le:NumberOfPG} and \ref{le:NumberOfPG2},
starting with the following inequality.
\begin{equation*} \sum_{G\in \pgi{S}{cN}} \hspace{-4mm} v_x(G) \le \sum_{q \in S} \pgr{S \setminus\{q\}}\enspace.
\end{equation*}\vspace{-10mm}

\end{theProof} 

We let $c_1 = 1.978993$ and $c_2 = 2.035802$.
Substituting these specific values of $c$ into Theorem \ref{th:triVSpgcc}, and using the bound $\tri{N} < 30^N$ of \cite{SS10}, we get
$\pgrci{N}{c_1}{1} = O^*\left(186.46^N\right)$ and $\pgrci{N}{c_2}{2} = O^*\left(180.20^N\right)$.

\begin{lemma}
\label{le:LowerX}
For every point set $S$ of $N$ points in the plane, we have:
\[\mbox{(i) Either } \pgri{S}{1} = O^*\left(186.46^N\right) \mbox{ or }\hv_{x,cN}(S,1) > N/186.46 \mbox{ (or both).}  \]
\[\mbox{(ii) Either } \pgri{S}{2} = O^*\left(180.20^N\right) \mbox{ or }\hv_{x,cN}(S,2) > N/180.20 \mbox{ (or both).}  \]
\end{lemma}
\begin{theProof}{\!\!}
We start by proving part (i). The proof goes along the same lines as the proof of Lemma \ref{le:LowerV02}. That is,
we use a charging scheme in which an $i$-ving $v=(p,G)$ is charged $a+2m/N-i$, where $a=(9-4c_1)/2$ and $m$ is the
number of edges in $G$.
As before, on average, every ving is charged at least $a$.
Each ving then moves its entire charge to its corresponding $x$-ving.
Since there are no 0-vings in this case, an $x$-ving is charged at most $2^{k-1}(2a+4m/N-k) - (a+2m/N)$ (the first
term is the charge of an $x$-ving according to the analysis of Lemma \ref{le:LowerV02}, and the second term is the
missing charge of the corresponding single 0-ving).
When $m/N=c_1$, the maximum charge is obtained when $k$ is either 7 or 8, and is then 123.5.

Next, we move charge from $x_7$-vings and $x_8$-vings to $x_3$-vings and $x_4$-vings,
in two steps, as in the proof of Lemma \ref{le:LowerV02}.
In the first step we move $20.8$ units of charge from every $x_7$-ving and $x_8$-ving
to the $x_3$-ving it reduces to, and in the second step we move 1.639 units from every
$x_7$-ving, $x_8$-ving, and the $x_3$-vings they reduce to, to the corresponding $x_4$-ving, just as in the
preceding analysis.
Initially, an $x_3$-ving gets a charge of 19.5, and after at most four $x_7$-vings and $x_8$-vings charge it
additional 20.8 units in the first step, every $x$-ving has a charge of at most $102.7 = 123.5-20.8=19.5 +4\cdot20.8$.
An $x_4$-ving starts with a charge of 35.5, and after at most forty $x$-vings charge it additional 1.639 units,
every $x$-ving has a charge of at most $101.061 = 102.7 - 1.639 > 35.5 +40\cdot1.639$.
Repeating the remaining part of the original analysis, we obtain that
either $\pgri{S}{1} = O^*\left(186.46^N\right)$ or
$\hv_{x,cN}(S,1) = \frac{aN}{101.061} = \frac{(9-4c)N}{202.122} > N/186.46$.

Part (ii), concerning the case where there are neither isolated vertices nor vertices of
degree 1, is proved in the same manner.  Since there are no 0-vings and no 1-vings in this
case, an $x$-ving is charged at most $2^{k-1}(2a+4m/N-k) - (a+2m/N) - k(a+2m/N-1)$
(we subtract the potential contributions of one 0-ving and $k$ 1-vings).
We set $a = 0.455955$, which, unlike the previous cases, does not satisfy $a=(9-4c_2)/2$.
Nevertheless, the maximum charge is still obtained when $k$ is either 7 or 8, and is then 102.307.
As before, we then move part of the charge to $x_3$-vings and $x_4$-vings, in two steps as before, where the charge
moved in the first step is $18.639$ and the charge moved in the second step is $1.508$.
An $x_3$-ving gets an initial charge
smaller than $9.111$, and then at most four $x_7$-vings and $x_8$-vings charge it an additional $18.639$, so every
$x$-ving has a charge of at most $83.668 = 102.307-18.639 > 9.111 +4\cdot 18.639$.
An $x_4$-ving gets an initial charge smaller than $21.804$, and after at most forty $x$-vings charge it an
additional $1.508$ in the second step, every $x$-ving has a charge of at most
$82.16 = 83.668 - 1.508 > 21.804 +40\cdot1.508$.
Repeating the remaining part of the original analysis, we obtain that
either $\pgri{S}{2} = O^*\left(180.20^N\right)$ or $\hv_{x,cN}(S,2) = \frac{aN}{82.16} > N/180.20$.
\end{theProof}

Recall that $\pgri{N}{i} = \max_{|S|=N}\pgri{S}{i}$.
By combining Lemma \ref{le:NumberOfPGno0} and Lemma \ref{le:LowerX}, we obtain the following bounds (the proof is essentially identical to the proof of Theorem \ref{th:pg2}).\vspace{2mm}

\begin{theorem} \label{th:pgi}
$\displaystyle \pgri{N}{1} = O^*\left(186.46^N\right)$ and $\pgri{N}{2} = O^*\left(180.20^N\right)$.
\end{theorem}

\noindent
{\bf Remark.} The above method can also be applied to bound $\pgri{N}{i}$ for $i=3,4,5$.
For $i \ge 6$ we have $\pgri{N}{i} = 0$, since every plane graph contains at least one vertex of degree smaller than
six. \vspace{2mm}

Since every connected graph has no isolated vertices, we obtain the following corollary.
\begin{corollary} \label{co:connected}
For every point set $S$ of $N$ points in the plane, the number of connected plane graphs that can be
embedded over $S$ is $O^*\left(186.46^N\right)$.
\end{corollary}
Although this is only a slight improvement over our bound of $O^*(187.53^N)$ on the total number of plane graphs,
this is nevertheless, as far as we know,
the first time that a bound on the number of connected plane graphs is asymptotically smaller than the bound on the
total number of plane graphs.
In a similar manner, since every bi-connected graph has no vertices of degree 0 or 1, we obtain
the following further improved bound.
\begin{corollary} \label{co:connected}
For every point set $S$ of $N$ points in the plane, the number of bi-connected plane graphs that
can be embedded over $S$ is $O^*\left(180.20^N\right)$.
\end{corollary}
The above method of cross-graph charging can also be used to obtain other properties of random plane graphs
(embedded over a fixed set of points).
For example, the following observation,  which is a variant of Lemma \ref{le:LowerV0},
applies our method to lower bound the expected number of $0$-vings, of 1-vings, and of 2-vings
in a graph uniformly chosen from $\pg{S}$. As already introduced in Section~\ref{ssec:direct},
we let the expected value of $v_i(G)$, for a graph
chosen uniformly at random from $\pg{S}$, be denoted as $\hv_i(S)$
\begin{lemma} \label{th:leaves}
For every set $S$ of $N$ points in the plane in general position,
\[ \hat{v}_0(S) \ge \frac{N}{4096}, \quad\hat{v}_1(S) \ge \frac{3N}{1024},
\quad \mbox{and} \quad \hv_2(S) \ge \frac{33N}{2048}. \]
\end{lemma}
\begin{theProof}{\!\!}
First, we notice the following bijective correspondence between $x$-vings and 0-vings.
An $x$-ving $v=(p,G)$ can be uniquely mapped to a 0-ving by removing from $G$ all the edges that are incident to $p$.
Similarly, a 0-ving can be uniquely mapped to an $x$-ving by connecting it to all the vertices that it can see.
This implies that, for every point set $S$, $\hat{v}_0(S) = \hat{v}_x(S)$, and thus,
the first part of the lemma is immediately implied by Lemma \ref{le:LowerV0imp}.

We next establish the second bound in the lemma.
We apply the same charging scheme as in Lemma \ref{le:LowerV0},
and then move the charge from $x$-vings to 1-vings in the following manner. Consider an $x_d$-ving $v=(p,G)$, and notice that exactly $d$ 1-vings correspond to $v$. We split the charge of $v$ evenly among these $d$ 1-vings.
In the proof of Lemma \ref{le:LowerV0}, it was shown that every $x_d$-ving
gets charged by exactly $2^{d-1}(14-d)$ units.
Therefore, each 1-ving that corresponds to $v$ is charged
$2^{d-1}(14-d)/d$, and no 1-ving can be charged more than once in this manner.
This expression is maximized at $d=12$, and its value is then  $1024/3$.
Thus, we can move the entire charge to 1-vings in a manner that guarantees that no 1-ving gets charged more than $1024/3$.
Since, as shown in Lemma \ref{le:LowerV0}, on average every ving is initially charged more than 1, we get that
$\hv_1(S) \ge \frac{3N}{1024}$, as asserted.

The bound for $\hv_2(S)$ is obtained in the exact same manner, except that in this case there are $d(d-1)/2$ 2-vings that correspond to $v$,
so the charge that each of them obtains is $2^d(14-d)/(d(d-1))$. The maximum value of $2048/33$ is attained at $d=12$,
implying the third claim of the lemma.
\end{theProof}

The proof of the lemma is based on the easy fact that every $x$-ving has degree $i\ge 2$.
The proof does not extend to $i \ge 3$, since there might be $x$-vings with no corresponding $i$-vings,
that is, of degree $2$, as depicted in Fig.~\ref{fi:0vings}(a).
Therefore, using the above charging scheme will not necessarily move the entire charge of the $x$-vings to the $i$-vings.
When $i$ is relatively small, we can partly overcome the above difficulty in the following manner.
\begin{lemma} \label{le:deg2}
For every set $S$ of $N$ points in the plane in general position,
$\displaystyle \hat{v}_2(S) + \hat{v}_3(S) \ge N/24$.
\end{lemma}
\begin{theProof}{\!\!}
As in the proof of Lemma \ref{th:leaves}, we move the entire charge to $x$-vings and then,
for each $x$-ving $v$, we evenly split the charge of $v$ between the 3-vings that correspond to it (if there are any).

Let us first consider an $x_d$-ving $v=(p,G)$, such that $d \ge 3$.
In this case, there are $\binom{d}{3} =d(d-1)(d-2)/6 $ 3-vings that correspond to $v$,
and thus, each of these 3-vings obtains a charge of $\displaystyle \frac{3\cdot 2^{d}(14-d)}{d(d-1)(d-2)}$.
This expression maximizes when $d$ is either 11 or 12, implying that such a 3-ving is charged at most $1024/55 < 18.62$.

As already noted, an $x$-ving cannot have degree smaller than two. Therefore, we are left only with the case where $d=2$.
In this case, $v$ gets a charge of 24, and we move this charge
to the only 2-ving that corresponds to $p$. Combining this with the case where $k \ge 3$, we notice that the entire charge was moved to 2-vings and 3-vings,
such that no ving gets charged more than 24, which implies the claim.
\end{theProof}

Various other lower bounds can be obtained in a similar manner.
The above bounds can be slightly improved by using the method presented in Section~\ref{ssec:firstImp}.

\section{Quasi-Plane Graphs} \label{sec:quasi}
In this section we show that our techniques can easily be extended to obtain singly-exponential
bounds for the number $k$-quasi-plane graphs, for $k=3,4$.

\paragraph{The number of 3-quasi-plane graphs.}
We use the notation given in the introduction. A 3-quasi-plane graph does not contain three pairwise crossing edges. Ackerman and Tardos \cite{AT07} proved that
such graphs have at most $6.5N-20$ edges, and that this is tight up to some additive constant. Using this result, we
can apply our method in a straightforward manner.
As before, we denote by $\qp{k}{S}$ the set of all $k$-quasi-plane graphs embedded (with straight edges)
on a fixed labeled set $S$ of $N$
points in the plane, and put $\qpg{k}{S}:= \left|\qp{k}{S}\right|$.
Moreover, we let $\qpg{k}{N}=\max_{|S|=N}\qpg{k}{S}$.

Given a $k$-quasi-plane graph $G \in \qp{k}{S}$, we say that a ving $v=(p,G)$ is an \emph{$x$-ving} if we cannot add
to $G$ any additional (straight)
edges that are adjacent to $p$ without violating the $k$-quasi-planarity property of $G$.
We denote by $v_i(G)$ the number of $i$-vings in $G$, and by $v_x(G)$ the number of $x$-vings in $G$ (as in the
previous scenarios).
The expected value of $v_x(G)$, for a graph chosen uniformly from $\qp{k}{S}$, is denoted as $\hv_x^k(S)$. More formally,
\[ \hv_x^k = \hv_x^k(S) := \ex{v_x^k(G)} = \frac{\sum_{G \in \qp{k}{S}}v_x(G)}{\qpg{k}{S}}\enspace. \]

\begin{lemma}\label{le:NumberOfQP}
For $N \geq 2$ and $k \ge 2$, let $\delta_{N}^k>0$ be a real number, such that
$\hat{v}_x^k(S) \geq \delta_{N}^k N$ holds for every set $S$ of $N$
points in the plane in general position. Then
$\displaystyle \qpg{k}{N} \leq \mbox{$ \frac{1}{\delta_{N}^k}$}\,\qpg{k}{N-1}$.
\end{lemma} \vspace{-2mm}

\begin{theProof}{\!\!}
We follow the proof of Lemma \ref{le:NumberOfPG}.
In doing so, we observe that, when inserting a new vertex $p$ into a $k$-quasi-plane graph $G$,
the set of vertices that $p$ can connect to (via straight edges) without violating $k$-quasi-planarity, is unique,
and the connecting edges are ``independent'', in the sense that adding any of them does not affect the eligibility
of the other edges to be added. This is because $k$-quasi-planarity can be violated by $k$ pairwise crossing edges,
and the newly added edges do not cross one another. The rest of the proof is identical to the earlier proof.
\end{theProof}
Next, we use the upper bound on the number of edges in a 3-quasi-plane graph to obtain a lower bound for $\delta_{N}^3$.
\begin{lemma}
\label{le:LowerV0k}
For every set $S$ of $N$ points in the plane, $\displaystyle \hv_x^3(S) \ge N/2^{26}$.
\end{lemma} \vspace{-1mm}
\begin{theProof}{\!\!}
We use a charging scheme where every $i$-ving $v=(p,G)$ will be charged $14-i$ units.
The sum of the charges of the vings in any fixed 3-quasi-plane graph $G \in \pg{S}$ is $\sum_i(14-i)v_i = 14\sum_iv_i - \sum_{i}iv_i = 14N - \sum_{i}iv_i$.
Since $G$ can have at most $6.5N-20$ edges, we have $\sum_{i}iv_i \le 13N-40$.
This implies that the total charge in any fixed graph is at least $14N - \sum_{i}iv_i \ge N+40$. Therefore, on average, every ving has a charge larger than 1.

Next, we move all of the charge to $x$-vings in the same manner as in Lemma \ref{le:LowerV0}.
As already observed, connecting a new edge to $p$
(while not violating the 3-quasi-plane property) does not affect the set of additional
edges that can be connected to $p$.
Consider the charge that an $x_d$-ving (a notation analogous to that used for plane graphs) $v=(p,G)$ can have.
By the observation just made, the number of $i$-vings that charge $v$ is exactly $\binom{d}{i}$, as before.
Therefore, $v$ is charged exactly
\[\sum_{i=0}^d\binom{d}{i}(14-i) = 14\sum_{i=0}^d\binom{d}{i}-\sum_{i=0}^d\binom{d}{i}i = 14\cdot 2^d - d\cdot 2^{d-1} = 2^{d-1}(28-d).\]

\noindent This expression maximizes when $d$ is either 26 or 27, and is then $2^{26}$.
Therefore, on average, a 3-quasi-plane graph on $S$ has more than $\frac{N}{2^{26}}$ $x$-vings.
\end{theProof}
By combining Lemmas \ref{le:NumberOfQP} and \ref{le:LowerV0k}, we obtain an upper bound on the number of 3-quasi-plane graphs.
As far as we know, this is the first exponential upper bound for $\qpg{3}{N}$.
\begin{theorem}
$\displaystyle \qpg{3}{N} \le 2^{26N}$.
\end{theorem}

{\bf Quasi-plane graphs with $\mathbf{\emph{k}\ge 4}$.}
Ackerman \cite{Ack09} proved that every 4-quasi-plane graph that is embedded over a set of $N$ points in the plane has at most $36N-72$ edges, even when the edges are not necessarily straight. This implies that $\qpg{4}{S}$ is also exponential in $N$. Specifically:
\begin{theorem} \label{th:qp4}
$\displaystyle \qpg{4}{N} \le 2^{145N}$.
\end{theorem}
\begin{theProof}{\!\!}
Since Lemma \ref{le:NumberOfQP} applies for every $k$, we only need to replace Lemma \ref{le:LowerV0k}.
We can derive the bound $\hv_d^4(S) \ge \frac{N}{2^{145}}$ by using the same analysis as in the proof of Lemma
\ref{le:LowerV0k}, except that an $i$-ving will be charged $73-i$ units.
In this case, the analysis implies that an $x_d$-ving is charged $2^{d-1}(146-d)$.
This expression is maximized for $d=144$ and $d=145$, and is then $2^{145}$.
\end{theProof}

A common conjecture (e.g., see \cite{Ack09,AT07,pach04}) is that every $k$-quasi-plane graph with $N$ vertices has at most  $c_k N$ edges,
where $c_k$ is a constant depending on $k$
(in fact, the conjecture is also made for the more general case where the edges are not necessarily straight).
Proving the conjecture will immediately imply that $\qpg{k}{N}$ is exponential in $N$ for every fixed $k$.
This consequence is easily obtained by adapting the proof of Theorem \ref{th:qp4},
and giving each $i$-ving a charge of $2c_k + 1 -i$.
Valtr \cite{Val97} proved that any $k$-quasi-plane graph with $N$ vertices has $O(N\log N)$ edges.
Combining this bound with the cross-graph charging technique only yields the superexponential bound $\qpg{k}{N} = (N/\log{N})^{O(N)}$.


\end{document}